\documentclass{article}
\usepackage[preprint]{spconf}
\usepackage{amsmath,graphicx}

\usepackage{multirow}
\usepackage{booktabs}
\usepackage{subcaption}
\usepackage{cite}

\title{Incremental Transfer Learning in Two-pass Information Bottleneck based Speaker Diarization System for Meetings}

\name{Nauman Dawalatabad$^{1}$, Srikanth Madikeri$^{2}$, C Chandra Sekhar$^{1}$, Hema A Murthy$^{1}$}
\address{$^{1}$Indian Institute of Technology Madras, India \\ $^{2}$Idiap Research Institute, Martigny, Switzerland}
\copyrightnotice{\small{\copyright\ IEEE 2019}}
\toappear{\small{To appear in {\it Proc.\ ICASSP 2019, May 12-17, 2019, Brighton, UK.}}}

\begin{document}
%
\maketitle
\begin{abstract}

The two-pass information bottleneck (TPIB) based speaker diarization system operates independently on different conversational recordings.
TPIB system does not consider previously learned speaker discriminative information while diarizing new conversations.
Hence, the real time factor (RTF) of TPIB system is high owing to the training time required for the artificial neural network (ANN). 
This paper attempts to improve the RTF of the TPIB system using an incremental transfer learning approach where the parameters learned by the ANN from other conversations are updated using current conversation rather than learning parameters from scratch. 
This reduces the RTF significantly.
The effectiveness of the proposed approach compared to the baseline IB and the TPIB systems is demonstrated on standard NIST and AMI conversational meeting datasets.
With a minor degradation in performance, the proposed system shows a significant improvement of 33.07\% and 24.45\% in RTF with respect to TPIB system on the NIST RT-04Eval and AMI-1 datasets, respectively.

\end{abstract}

\section{Introduction}

Speaker diarization systems determine \textit{{``who spoke when?"}} in a conversational recording \cite{xavier12-review,tranter06-ovr}.   
Diarization has applications in telephone conversations, meetings, broadcast news, and TV shows.  
Diarizing conversational meetings is challenging owing to the dynamic nature of speaker switching. 
The diarization output is often used as a front-end in speech applications like automatic speech recognition  \cite{Dimitriadis2017} and speaker linking in a large corpus \cite{sturim2016speaker,srikanth16-linking}. 
Agglomerative hierarchical clustering is the most used approach for diarization, where short segments of speech are clustered in a bottom-up manner.
Speaker diarization systems based on hidden Markov model/ Gaussian mixture model (HMM/GMM) \cite{ajmera03-robust} and information bottleneck (IB) approach \cite{deepu09-ib} are the most popular.

Real Time Factor (RTF) is defined as the ratio of the time taken by an algorithm to the duration of the input.
IB based diarization systems are significantly fast and are suitable for real time applications \cite{deepu09-ib,srikanth15-improveRuntime}.
To enable better speaker discrimination, there has been numerous efforts in finding the deep neural network (DNN) based speaker discriminative embeddings \cite{yella14-ann,le17-tripletloss,Bredin2017TristouNetTL,garcia17-dnnembedding,sd-lstm,xvect,jati2018:npc}.
Different network architectures and/or loss functions are used to obtain speaker discriminative representations.
However, in all cases, the DNN needs to be trained on huge amounts of labeled/unlabelled data. 
The two-pass IB based speaker diarization (TPIB) \cite{dawalatabad16-tisd} system is a recently proposed unsupervised approach that does not use any separate training data to learn speaker discriminative characteristics.
In this approach, the speaker discriminating features are automatically learned from the current conversational recording to be diarized.
Although the error rates reported by TPIB system are better than the IB system, the RTF is high mainly due to the increased time for training the artificial neural network (ANN). 

In this paper, we attempt to improve the RTF of the TPIB system. 
All the unsupervised methods (including the TPIB system)  operate independently on different audio recordings under consideration.
The information learned during diarizing one recording can be used to improve the diarization on another recording.
The system should remember the previous speaker discriminating knowledge, learn new information and then transfer this information incrementally for diarizing other audio recordings.
This {\it ``Remember--Learn--Transfer''} approach is the natural way for humans. 
We propose to use transfer learning across recordings in an incremental fashion to learn speaker discriminating features continuously.

The paper is organized as follows.
Section \ref{sec:tisd} briefly explains the IB and TPIB systems.
In Section \ref{sec:proposed} we describe the proposed system.
Section \ref{sec:res} presents the results of all systems on different datasets. 
Finally, Section \ref{sec:con} concludes the paper.

\begin{figure}[t]
		\includegraphics[scale=0.135]{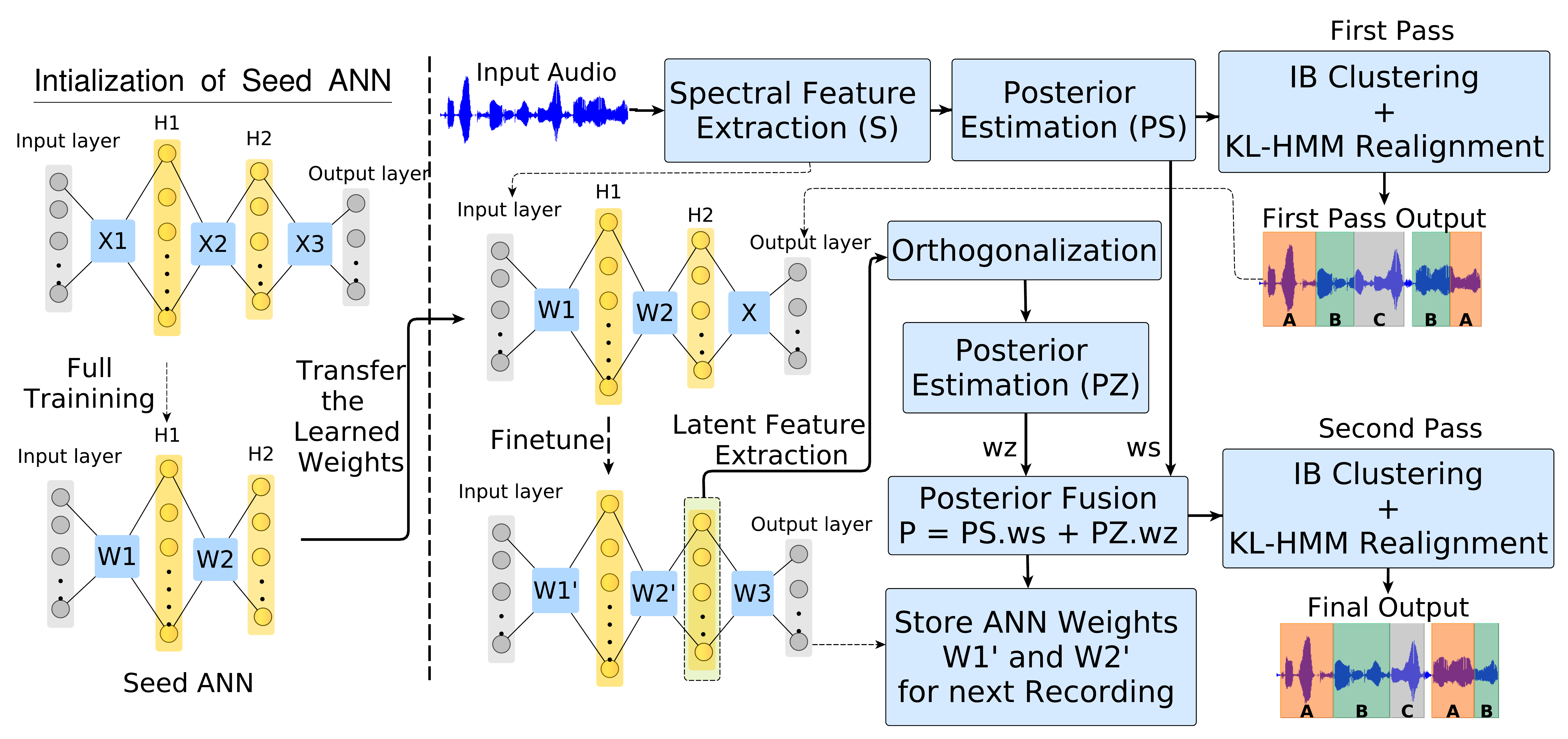}  
        \caption{{\it Block diagram of the proposed system (TPIB-ITL).}} 
        \label{fig:tpib-tl}
\end{figure}

\vspace{-0.1cm}

\section{Information Botlleneck based Systems}
\label{sec:tisd}
\vspace{-0.1cm}
This section briefly describes the IB and TPIB systems.
\vspace{-0.3cm}
\subsection{Baseline Information Bottleneck System}
\label{subsec:ib}
\vspace{-0.1cm}

In agglomerative information bottleneck algorithm \cite{deepu09-ib}, short segments $\mathbf{X}$ of speech are clustered in a bottom-up manner. 
A GMM is modeled using $\mathbf{X}$ where the GMM components are denoted by $\mathbf{Y}$, and represent relevant information. 
The IB based approach clusters the set of segments $\mathbf{X}$ into a set of clusters $\mathbf{C}$ such that most of the relevant information about $\mathbf{Y}$ is captured.
The objective function $\mathcal{F}$ is given by
\begin{equation}
\label{eq:obj}
\mathcal{F} =  \mathrm{I}( \mathbf{Y}, \mathbf{C}) - \mathbf{\frac{1}{\beta}} \mathrm{I}( \mathbf{C}, \mathbf{X})
\end{equation}
where $ \mathrm{I(\cdot)}$ denotes the mutual information and \( \mathbf{\beta} \) is a Lagrange multiplier. 
The normalized mutual information (NMI) \cite{deepu09-ib} is used as a criterion for terminating the clustering process.

\subsection{Two-pass IB based Diarization System}
\label{subsec:tisd}
Two-pass IB (TPIB) based speaker diarization system \cite{dawalatabad16-tisd} consists of 2 passes of the IB diarization as given below. 

\noindent - {\it \textbf {First pass}}: 
In the first pass, standard IB based diarization is performed followed by Kullback-Leibler hidden Markov model (KL-HMM) based realignment \cite{deepu09-klRealign}.

\noindent - {\it \textbf {ANN Training, Feature Extraction and Orthogonalization}}:  
ANN initialized with random weights is trained on the output boundary labels and the spectral features obtained in the first pass.
Latent features are extracted from the penultimate layer of the trained ANN. 
Principal component analysis (PCA) is then applied to whiten the features.

\noindent - {\it \textbf {Second pass}}: The latent features are used along with the spectral features in the second pass of IB clustering which is then followed by a KL-HMM based realignment.


\section{TPIB with Incremental  Transfer Learning}
\label{sec:proposed}

Transfer learning is the improvement of learning in a new task through the transfer of knowledge from a related task that has already been learned \cite{Olivas:2009}.
In TPIB system, the ANN learns the parameters from scratch for each conversation independent of the other.
It is the primary bottleneck in terms of run time.
We attempt to solve this problem by retaining the knowledge learned by the model from other audio recordings. 
New discriminative information from the current recording is used to learn more speaker discriminative features.
As shown in Fig. \ref{fig:tpib-tl}, the following steps are involved in the proposed TPIB system with incremental transfer learning (TPIB-ITL). 

\noindent {\it - \textbf{Step 0 - Training seedANN:}}
A $seedANN$ is trained from the first audio to be diarized by the system.
We first perform IB based diarization followed by KL-HMM based realignment to get the relative speaker labels.
We intialize all layers' weights ($X1$, $X2$,  and $X3$) of $seedANN$ using Xavier intialization technique as described by Glorot and Bengio in \cite{glorot2010understanding}. 
The obtained speaker labels and the spectral features are used to train the $seedANN$.
This $seedANN$ is used for the subsequent recordings to be diarized.
It is important to note that no separate data is used for training the $seedANN$.

\noindent {\it - \textbf{Step 1 - First pass:}}
In the first pass, we perform IB based diarization that is followed by KL-HMM based realignment.

\noindent {\it - \textbf{Step 2 - Transfer of knowledge and fine-tuning:}}
Unlike random initialization of ANN weights in TPIB, we initialize the parameters of the current ANN with the parameters of the $seedANN$.
However, the number of speaker labels in the first-pass output of the current recording can be different from the number of output neurons in the $seedANN$.
To handle this mismatch, the parameters $W1$ and $W2$ (as shown in Fig. \ref{fig:tpib-tl}) are initialized from the $seedANN$.
The final layer weights alone are initialized using Xavier initialization ($X$).
After initialization, fine-tuning of the ANN is performed for a small number of epochs. 
Fine-tuning is performed using the first pass output labels and the spectral features of the current recording to be diarized.
The fine-tuned ANN weights ($W1^{'}$ and $W2^{'}$) are stored as $ANN^{*}$.

\noindent {\it - \textbf{Step 3 - Latent Feature Extraction and Orthogonalization:}}
Once the ANN is fine-tuned for the current recording, the output of the penultimate layer is used as the latent feature (LF) representation.
The output features are subjected to PCA for orthogonalization.

\noindent {\it - \textbf{Step 4 - Posterior Merging:}}
The latent space features (LSF) are merged with the spectral features \cite{deepu09-ib} as, 
\begin{equation} 
\label{eq:fuse}
P(\mathbf{y|f_t^{s}, f_t^{z}}) = P(\mathbf{y|f_t^{s}}).w_{s} + P(\mathbf{y|f_t^{z}}).w_{z}
\end{equation}
where $\mathbf{f_t^s}$ and $\mathbf{f_t^z}$ are feature vectors at time $t$ from spectral feature stream $\mathbf{s}$ and latent feature stream $\mathbf{z}$, respectively.
Here, $w_{s}$ and ${w_{z}}$ are the weights assigned to the feature streams $\mathbf{s}$ and $\mathbf{z}$ respectively, such that ${w_{s}}$ + ${w_{z}}$ = 1.

\noindent {\it - \textbf{Step 5 - Second pass:}}
Finally, the second pass of IB clustering is performed on $P(\mathbf{y|f_t^{s}, f_t^{z}})$ which is followed by KL-HMM based realignment.
This is the final diarization output.

For every new recording, the parameters of the ANN is initialized with the $ANN^{*}$.
Notice that $ANN^{*}$ gets updated after every fine-tuning step ({\it Step 2}) for each recording.

\vspace{-0.2cm}
\subsection{Remember-Learn-Transfer}
\vspace{-0.1cm}

The neural network primarily learns to discriminate between speakers in a recording.   
The characteristics can therefore be similar across the recordings.
The discrimination learned by the ANN on one conversational recording can be useful for training an ANN on another recording.
This is missing in the TPIB framework where the already learned speaker discriminative information is never utilized for training any future model.
Transfer learning enables the new model to start learning from a better point than starting from random initialization, which ultimately leads to faster convergence.

The fine-tuning of ANN is essential for the current recording as we are interested in discriminating the speakers in that recording. 
Moreover, this also helps to get rid of problems of unseen speakers (or classes).
The fine-tuning is performed on every new recording, which can lead to additional speaker discriminative information.
Hence, the ANN model gets better at discrimination as it continuously learns from previously diarized conversations.

\vspace{-0.1cm}

\section{Experimental Setup and Results}
\label{sec:res}
\vspace{-0.15cm}
This section explains the datasets, experimental setup and results obtained.

\begin{table}[t]
\centering
\caption{AMI meeting datasets.}
\vspace{-0.2cm}
\label{tab:ami}
\resizebox{0.4\textwidth}{!}{
\begin{tabular}{@{}c|c@{}}
\toprule
\textbf{AMI-1} & \begin{tabular}[c]{@{}c@{}}ES2008c, ES2013a, ES2013c, ES2014d, ES2015a,\\ IS1001c, IS1007a, IS1008c, IS1008d, IS1009c\end{tabular} \\ \midrule
\textbf{AMI-2}          & \begin{tabular}[c]{@{}c@{}}ES2010b, ES2013b, ES2014c, ES2015b, ES2015c,\\ IS1004b, IS1006c, IS1007c, IS1008a, IS1009d\end{tabular}          \\ \bottomrule
\end{tabular}}
\vspace{-0.2cm}
\end{table}

\vspace{-0.2cm}
\subsection{Dataset, Features and Evaluation Measure}
\vspace{-0.1cm}
Experiments are performed on standard NIST-RT (RT-04Dev, RT-04Eval, RT-05Eval) \cite{nist} datasets and  on the subsets of  Augmented Multi-Party Interaction (AMI) corpus \cite{ami}. 
The list of meeting IDs from AMI datasets recorded at Idiap (IS) and Edinburgh (ES) is given in Table \ref{tab:ami}.
These are the same AMI meetings selected randomly in \cite{dawalatabad16-tisd} for evaluation. 
The number of speakers in the NIST dataset ranges from 3 to 8 whereas there are four speakers in each of the AMI recordings.
Most recordings have a different set of speakers. 
RT-04Dev was used as the development dataset and remaining datasets were used for the testing purpose. 

Mel frequency cepstral coefficients (MFCC) of 19 dimensions extracted with 10 ms shift from 26 filterbanks are used as input spectral features.
Diarization error rate (DER) is the sum of missed speech (MS), false alarm (FA) and speaker error rate (SER).
The MS and FA depend on errors in speech activity detection whereas the SER is due to the speaker mismatch.
As the focus of the paper is primarily on clustering, speech/non-speech hypotheses were obtained from the ground truth and SER is used as the evaluation metric. 
As the proposed system does not use any separate training data (labelled/unlabelled) to obtain speaker discriminative features, we compare the performance of the proposed system to the baseline IB and the TPIB systems. 
We kept a forgiveness collar of 0.25 sec and included overlapped speech in our evaluations. 

\vspace{-0.25cm}
\subsection{Models and Experimental Setup}
\vspace{-0.1cm}
A  multi-layer feed-forward neural network (MLFFNN) with 2 hidden layers was used for both TPIB and TPIB-ITL systems.
The first layer has 30 neurons with $\tanh$ activation function and the second layer has 16 neurons with linear function.
We use TensorFlow's implementation of stochastic gradient descent with cross-entropy loss function to train/fine-tune the ANN models.

All the weights were initialized using Xavier initialization.
The early stopping criterion based on cross-entropy error was set to achieve the best results on the development set in terms of both, SER and RTF.
This implementation of TPIB system gave lower RTF for TPIB than that reported in  \cite{dawalatabad16-tisd}. 

The results reported in Table \ref{tab:ser} for TPIB-ITL are when the system is executed in the standard chronological order of the meeting Ids within the dataset.
The sequence of the datasets used is also the same as shown (left to right) in Table \ref{tab:ser}.
This standard sequence is maintained just for reproducibility of the results.
However, similar trends were observed when the TPIB-ITL system was run on 10 different random permutations of the meeting sequences with different recording for $seedANN$. 
As we will show in the next section, the order of meetings does not influence the overall performance of TPIB-ITL.
As the system uses information from previous recordings, an alternative experiment was also conducted.   
Here, the incremental transfer learning is performed only on development data. 
The parameters of the ANN for each test conversation are then fine-tuned from the trained ANN independent of the other conversations. 
The trained ANN is used for all the test recordings independently.
The SERs for this experiment is given in TPIB-ITL (Dev.) (i.e., last row of Table \ref{tab:ser}). 
The number of fine-tuning epochs was set to 50 for both TPIB-ITL and TPIB-ITL (Dev.).

We use open source IB toolkit \cite{deepu12:diartk} in all the experiments.
The values of NMI and $\beta$ were set to 0.4 and 10, respectively.
All RTFs are calculated on 2.6 GHz CPU with 2 threads.
The reported RTFs are calculated by averaging the RTFs across 10 independent runs. 
All the results reported in this paper are based on the best performing parameters tuned on the development dataset.

\begin{table}[t]
\centering
\caption{Speaker Error Rate (SER) on different systems are mentioned.
The feature fusing weights are mentioned in parentheses. 
Avg. denotes the average SER over all fusing weights combination.
Best SER on both systems for each dataset is indicated in bold font.
}
\vspace{-0.2cm}
\label{tab:ser}
\resizebox{0.47\textwidth}{!}{
\begin{tabular}{@{}clccccc@{}}
\toprule
\multicolumn{1}{l}{\multirow{2}{*}{System}} & \multirow{2}{*}{Feature(s)} & Dev. Set & \multicolumn{4}{c}{Test Sets}          \\   \cmidrule(l){4-7} \cmidrule(l){3-3}
\multicolumn{1}{l}{}                        &                             & RT-04Dev & RT-04Eval & RT-05Eval & AMI-1 & AMI-2 \\ \midrule
 IB                            & MFCC                        & 15.1            	& 13.5      & 16.4      & 17.9  & 23.5  \\  \midrule
\multirow{3}{*}{TPIB }                       & LSF                      & 15.1         	& 11.6      & 14.2      & 17.5  & 21.3  \\
                                            & MFCC+LSF (0.8, 0.2)      & 13.1     & 12.5      & 16.6      & 16.4  & 22.7  \\
                                            & MFCC+LSF (Avg.)           & 14.9     	& 12.6      		& 15.3      & 17.8  & 22.4  \\  \midrule 
\multicolumn{7}{c}{Proposed System}                                                    \\ \midrule                                        
\multirow{3}{*}{TPIB-ITL}                    & LSF                     & 15.5     	& 12.5      		& 15.1      			&  {\bf 17.5}  & 22    \\
                                            & MFCC+LSF (0.1, 0.9)        & 15.2     & {\bf 12.2}      		&  {\bf 15}        & 18    &  {\bf 21.2}  \\
                                            & MFCC+LSF (Avg.)             & 15.8     & 12.5      		& 15.4      & 17.8  & 21.9  \\ \midrule \midrule
\multirow{3}{*}{TPIB-ITL (Dev.)}                    & LSF                     & 15.5     	& 12.9      		& {\bf 14.8}      			&  17.5  & 22.1    \\
                                            & MFCC+LSF (0.1, 0.9)        & 15.2     & {\bf 12.5}      		&  15        & {\bf 17.5}    &  {\bf 22}  \\
                                            & MFCC+LSF (Avg.)             & 15.8     & 13.3      		& 15.6      & 17.8  & 22.5  \\ 
                                            
\bottomrule
\end{tabular}}
\vspace{-0.35cm}
\end{table}

\subsection{Results and Discussion}
\vspace{-0.1cm}
The SERs for different approaches to diarization are given in Table \ref{tab:ser}.
It can be seen that both the proposed system (TPIB-ITL) and the TPIB system outperform the baseline IB system in most cases.
The SERs of TPIB and TPIB-ITL systems are comparable. 
The TPIB-ITL system showed an absolute improvement of 2.3\% with respect to the baseline IB system on AMI-2 dataset. 
Though the best feature combination tuned on development set gives better SER than the IB system, we observed that it may not always show improvement over LSF.
This happens when the extracted LSF itself is better at speaker discrimination than the combination.
Hence to check the overall behaviour of the system under different feature fusing weights, we also provide the averaged SER across all the feature fusing weights for each dataset.
It gives a measure of robustness of the system to the feature fusing weights.
It can be seen from Table \ref{tab:ser} that the average SER of the proposed system show absolute 1.6\% improvement on AMI-2 dataset.
This is significant, as it is calculated over the average of all possible feature fusing weight combinations.
It confirms the robustness of the TPIB-ITL system to the feature fusing weight combinations.

The SER on development data for TPIB-ITL is slightly higher than other systems.
This is expected as the ANN has learned speaker discrimination from only a few recordings. 
Its performance is better on the test data, as incremental transfer learning happens. 
The improvement in SER can be observed on all test datasets. 
Average SER (avg.) of 21.9\% is observed for TPIB-ITL system on AMI-2 (last dataset in sequence) whereas it is 22.4\% for TPIB system.
This confirms that the ANN model in TPIB-ITL gets better over time.

To ensure that the results are not biased towards ordering of meetings, experiments were conducted for random permutations with random $seedANN$.
The overall trend in the result was observed to be similar.
For example, the average SER for TPIB-ITL over all the feature fusing weights on each dataset for one of the random permutations of the meeting sequence different from the sequence shown in Table \ref{tab:ser} is 
RT-04Dev=15.5, RT-04Eval=12.6, RT-05Eval=14.9, AMI-1=17.7, AMI-2=22.2.
This confirms that the sequence of recordings does not influence the overall performance of the proposed system.

\begin{table}[t]
\centering
\caption{The RTF on different systems for different datasets are mentioned.
Impr. denotes the relative  improvement in RTF with respect to TPIB system.
The observed deviation in RTFs  is between 0.002--0.004 on NIST datasets and  0.003--0.007 on AMI datasets.}
\label{tab:rtf}
\vspace{-0.25cm}
\resizebox{0.45\textwidth}{!}{
\begin{tabular}{cccccc}
\toprule
Sys/Dataset     & RT-04Dev & RT-04Eval & RT-05Eval & AMI-1 & AMI-2 \\ \midrule
IB         & 0.070      & 0.081     & 0.086      & 0.241 & 0.304  \\
TPIB       & 0.248      & 0.257     & 0.254     & 0.642 & 0.740 \\ \midrule \midrule
TPIB-ITL    & 0.175    & 0.172     & 0.180     & 0.485 & 0.605 \\ 
Impr. (\%) & 29.44    & \textbf{33.07}     & 29.13     & 24.45 & 18.24 \\ 
\bottomrule
\end{tabular}}
\vspace{-0.5cm}
\end{table}

The results where the training of ANN is used incrementally in TPIB-ITL using only development data is given in TPIB-ITL(Dev.)  (last row of the Table \ref{tab:ser}).
The system outperforms the baseline IB system and is comparable to TPIB system.
Notice that the system does not use any labeled information from the development dataset (RT-04Dev). 
Moreover, it should also be noted that the RT-04Dev is a very small dataset comprising of only 8 meeting recordings.

A significant improvement in RTF is obtained over the TPIB system.
It can be seen from Table \ref{tab:rtf} that the best case improvement of 33.07\% in RTF is observed on RT-04Eval dataset for TPIB-ITL system.
The RTFs for all NIST-RT datasets are in a similar range. 
As the meeting duration in AMI datasets is longer than that in NIST-RT datasets, the RTF of the IB system itself is more for AMI datasets.
Nevertheless, we observe improvements in RTF of 24.45\% and 18.24\% for AMI-1 and AMI-2 datasets, respectively.  
Though TPIB-ITL system seems inherently sequential, it can be easily run in parallel on different batches of recordings.      
The IB system showed an average (over all datasets) error of 2.2  in estimating a correct number of speakers, whereas the TPIB and TPIB-ITL showed a similar average error of 1.15 speakers.

\begin{figure}[t]
    \centering
   \hspace{-0.9cm}
    \begin{subfigure}[b]{0.2\textwidth}
    \centering
		\includegraphics[scale=0.42]{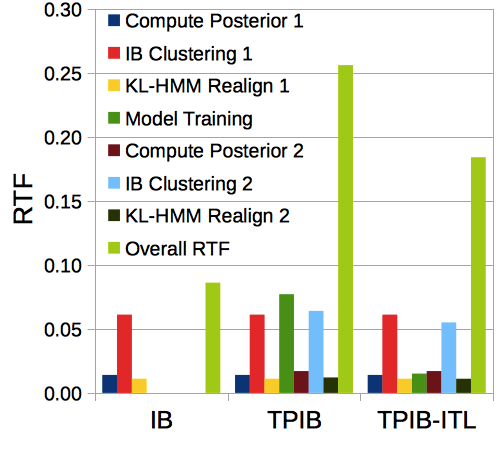}        
        \caption{RT-05Eval Dataset}
    \end{subfigure}%
     ~~~~~~~
    \begin{subfigure}[b]{0.2\textwidth}
    \centering
		\includegraphics[scale=0.42]{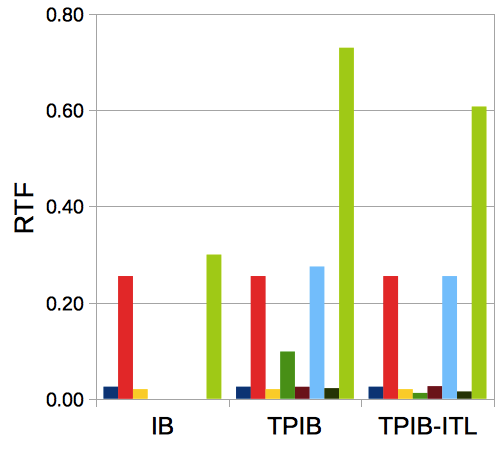}        
        \caption{AMI-2 Dataset}
    \end{subfigure}
     \vspace{-0.2cm}
      \caption{RTFs of individual modules for all systems.}
    \label{fig:module_rtf}
    \vspace{-0.45cm}
\end{figure}

Fig. \ref{fig:module_rtf} shows module-wise RTF for all the systems on two datasets; NIST RT-05Eval and AMI-2. 
Other respective NIST and AMI datasets also show a similar trend.
As the orthogonalization step takes negligible time, we do not report it. 
From Fig. \ref{fig:module_rtf}, it can be seen that considerable time is consumed in the ANN model training stage which is reduced significantly by TPIB-ITL system on both NIST and AMI datasets.


\vspace{-0.24cm}
\section{Conclusion} 
\label{sec:con}
\vspace{-0.2cm}
The proposed TPIB-ITL system uses {\it ``Remember-Learn-Transfer''} principle to incrementally learn speaker discriminative characteristics.
The ANN model gets better at speaker discrimination over time even with a small number of fine-tuning epochs.
With a minor degradation in performance, the proposed system show a significant improvement of 33.07\% and 24.45\% in RTF on RT-04Eval and AMI-1 datasets over TPIB system, respectively.
Since  the proposed system follows the TPIB framework, it also inherits the advantages of the TPIB framework, i.e., (i)  No separate training data is needed, and (ii) It incorporates recording-specific speaker discriminative features during the diarization process. 

\bibliographystyle{IEEEbib}
\bibliography{diary_tl}

\end{document}